%% file: cieran.tex
\documentclass[sigconf]{acmart} 
\usepackage{algorithm}
\usepackage{algpseudocode}
\usepackage{stfloats}

\algrenewcommand\algorithmicrequire{\textbf{Input:}}
\algrenewcommand\algorithmicensure{\textbf{Output:}}

\definecolor{keywordcolor}{rgb}{0.13,0.29,0.53} 
\definecolor{classcolor}{rgb}{0.31,0.60,0.02} 
\definecolor{methodcolor}{rgb}{0.63,0.81,0.181} 
\definecolor{paramcolor}{rgb}{0.6,0.4,0.8} 

\definecolor{teachcolor}{RGB}{0,0,0} 
\definecolor{outputcolor}{RGB}{0,0,0} 
\definecolor{initcolor}{RGB}{0,0,0} 

\newcommand{\add}[1]{#1}
\newcommand{\del}[1]{}

\AtBeginDocument{%
  }

\copyrightyear{2024}
\acmYear{2024}
\setcopyright{acmlicensed}\acmConference[CHI '24]{Proceedings of the CHI Conference on Human Factors in Computing Systems}{May 11--16, 2024}{Honolulu, HI, USA}
\acmBooktitle{Proceedings of the CHI Conference on Human Factors in Computing Systems (CHI '24), May 11--16, 2024, Honolulu, HI, USA}
\acmDOI{10.1145/3613904.3642903}
\acmISBN{979-8-4007-0330-0/24/05}

\begin{document}

\newcommand{\rev}[1]{\textcolor{black}{#1}}

\title[Cieran: Designing Sequential Colormaps via In-Situ Active Preference Learning]{Cieran: Designing Sequential Colormaps \\via In-Situ Active Preference Learning}

\author{Matt-Heun Hong}
\affiliation{%
  \department{Department of Computer Science}
  \institution{University of North Carolina\\ at Chapel Hill}
  \city{Chapel Hill, NC}
  \country{USA}}

\author{Zachary N. Sunberg}
\affiliation{%
  \department{Smead Aerospace Engineering Sciences Department}
  \institution{University of Colorado Boulder}
  \city{Boulder, CO}
  \country{USA}
}

\author{Danielle Albers Szafir}
\affiliation{%
  \department{Department of Computer Science}
  \institution{University of North Carolina\\ at Chapel Hill}
  \city{Chapel Hill, NC}
  \country{USA}}

\renewcommand{\shortauthors}{Hong, et al.}

\begin{abstract}
Quality colormaps can help communicate important data patterns. However, finding an aesthetically pleasing colormap that looks ``just right'' for a given scenario requires significant design and technical expertise. We introduce Cieran, a tool that allows any data analyst to rapidly find quality colormaps while designing charts within Jupyter Notebooks. Our system 
employs an active preference learning paradigm to rank expert-designed colormaps and create new ones from pairwise comparisons, allowing analysts who are novices in color design to tailor colormaps to their data context. We accomplish this by treating colormap design as a path planning problem through the CIELAB colorspace with a context-specific reward model. In an evaluation with twelve scientists, we found that Cieran effectively modeled user preferences to rank colormaps and leveraged this model to create new quality designs. Our work shows the potential of active preference learning for supporting efficient visualization design optimization.

\end{abstract}

\ccsdesc[300]{Human-centered computing~Interactive systems and tools}
\ccsdesc[500]{Human-centered computing~Visualization systems and tools}
\ccsdesc[300]{Computing methodologies~Q-learning}
\ccsdesc[300]{Computing methodologies~Learning from demonstrations}
\ccsdesc[500]{Computing methodologies~Learning from implicit feedback}
\ccsdesc[500]{Computing methodologies~Active learning settings}

\keywords{visualization, colormaps, design optimization, preference learning}



\begin{teaserfigure}
  \centering
  \includegraphics[width=\textwidth]{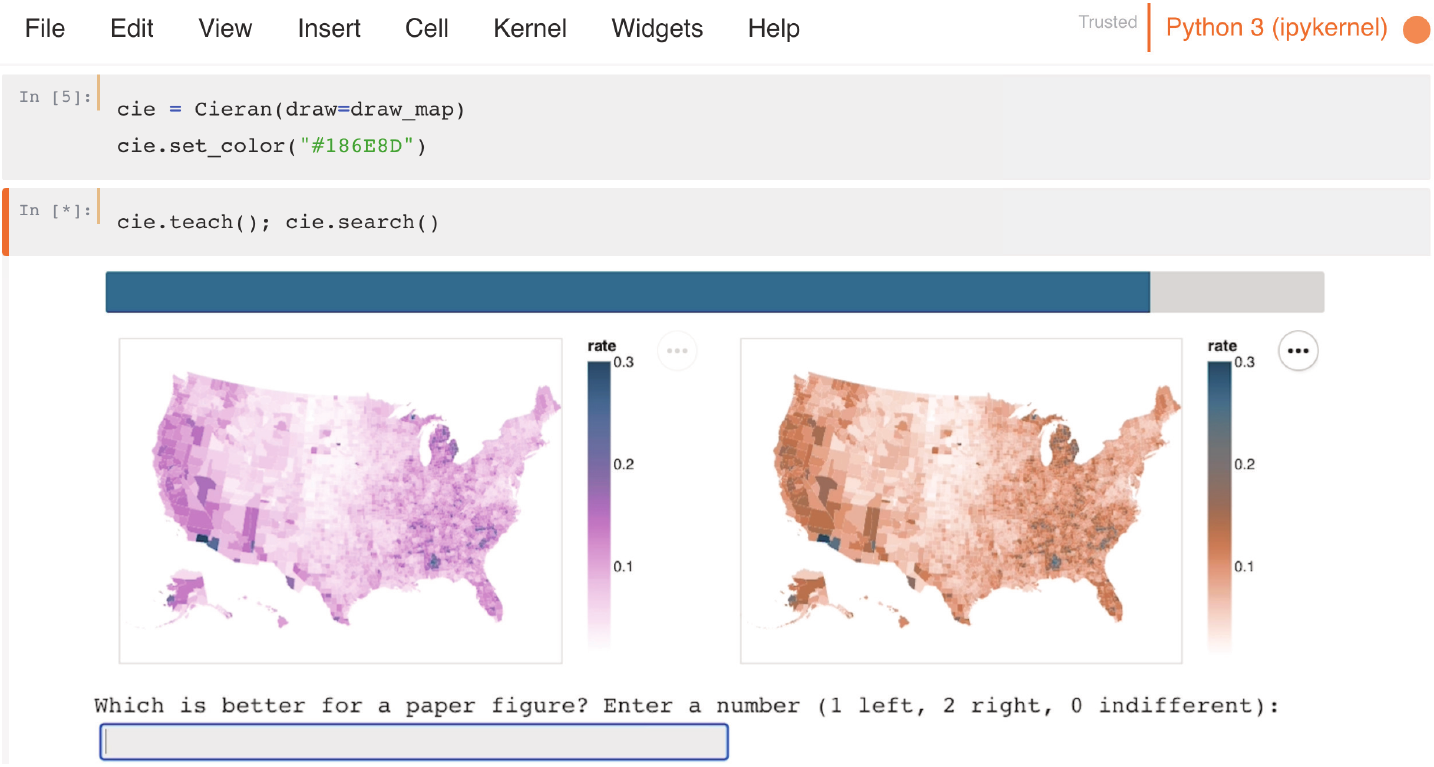}
  \caption{\textbf{Cieran's preference learning interface used with Altair \cite{altair}.} \textmd{Cieran supports efficient colormap selection within an analyst's workflow through integration with Jupyter Notebooks. 
  Cieran first interpolates example colormaps through a chosen color (e.g., \texttt{\#186E8D}, a teal blue shown in the progress bar). 
  People iteratively input preferences to Cieran by making value judgements across pairs of examples, and Cieran uses the preference data to induce a context-specific model of aesthetic utility. This model used to rank and create colormaps. }}
  \Description[A screenshot of using Cieran within a Jupyter notebook.]{A screenshot of using Cieran within a Jupyter notebook. The notebook displays two cells. In the first cell, the user instantiates Cieran by passing in a function called draw_map as a parameter. The user then sets the seed color. In the second cell, the user starts the preference learning process. The output of the second cell displays two of the same choropleth maps, differing only in the colormaps applied to the visualization. There is a text input component below the two maps where the user will input their choice.}
  \label{fig:interface}
\end{teaserfigure}

\maketitle

\section{Introduction}
\label{introduction}
\input{tex/introduction}

\begin{figure*}[t]
  \centering
  \includegraphics[width=1\textwidth]{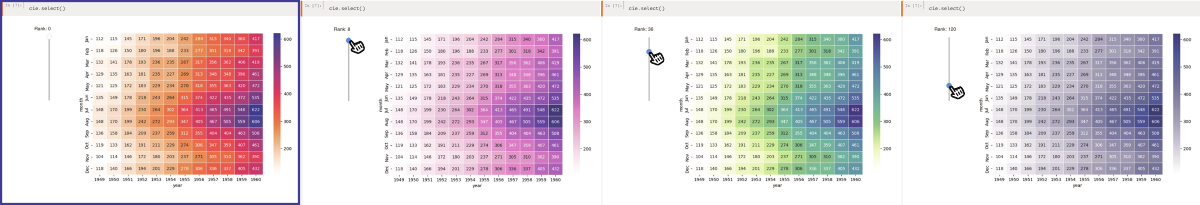}
  \caption{\textbf{Cieran allows analysts to quickly design a colormap for a visualization.} \textmd{Cieran creates a new colormap (left) and ranks existing expert-designed colormaps (the three sorted options to the right). After training Cieran, the user makes a selection using a slider widget, with the new and the most useful example colormaps sorted to the top. This gives users the final agency over the colormap design. 
  }}
  \Description[Cieran's colormap selection tool.]{A screenshot of the final selection tool of Cieran. The selection tool allows users to explore a variety of ranked colormap options using a vertical slider.}
  \label{fig:options}
\end{figure*}

\section{Background}
\label{background}
\input{tex/background}

\label{cieran}
\input{tex/cieran}

\section{System Evaluation}
\label{evaluation}
\input{tex/evaluation}

\section{Discussion}
\label{discussion}
\input{tex/discussion}

\begin{figure}[b]
  \centering
  \includegraphics[width=1.0\columnwidth]{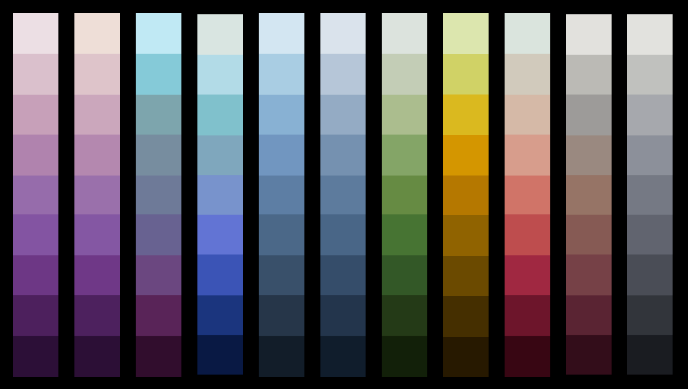}
  \caption{\textbf{A set of novel colormaps created by Cieran.} \textmd{Each colormap above was found by Cieran using Algorithm \ref{q-learning} according to a study participant's preferences, and voted by the participant to be the best across the four options shown to them (Section \ref{ranking-task}).}}
  \Description{Eleven different new created colormaps arranged horizontally.}
  \label{fig:colormaps}
\end{figure}

\section{Future Work}
\label{limitations}
\input{tex/limitations}

\section{Conclusion}
To help domain experts customize colormaps optimized to their target context, our system Cieran leverages active preference learning to model the aesthetic utility of expert-designed colormap curves to rank hundreds of design options available to users (\S\ref{sec:ranking}).
Cieran is also able to rapidly find novel colormaps (\S\ref{sec:searching}) that are highly useful (\S\ref{sec:quant}). While grounded in expert examples, our system also processes colormaps to meet known best practices in colormap design.
Specifically, our system leverages a preference-based reward learning algorithm, reinforcement learning, and a Jupyter Widgets interface for Python users to optimize their charts \textit{in situ}, regardless of which plotting library was used.
Our evaluation demonstrated the efficacy of this paradigm, which entailed asking only a simple question like ``Which chart do you like better?'' for optimizing the look-and-feel of a chart across the complex space of colormap design.
We hope that this work can inspire the development of future design automation tools for data visualization.

\begin{acks}
This work was supported by NSF awards \#1764089, \#1764092 \& \#2046725.
\end{acks}

\balance
\bibliographystyle{ACM-Reference-Format}
\bibliography{cieran}
\clearpage
\appendix


\section{Uniform Parametrization of Curves by Arc Length}
\label{appendix-A}

A curve can be defined as a parametric function $C(t)$, where $t$ is parameter point in the continuous interval $[0, 1]$. Our goal is to find a discrete and increasing parameter set $\{t^*_0, \cdots, t^*_N \}$ such that the distance between any two consecutive coordinates $D(C(t^*_i), C(t^*_{i+1}))$ is the same.

The following algorithm was also used to design perceptually uniform \texttt{matplotlib} colormaps (magma, inferno, plasma, and viridis) \cite{viscm}. 
However, these colormaps utilize simple Euclidean distances between sampled colors in the CIELAB colorspace. In contrast, Cieran uses the $\Delta E_{2000}$ perceptual distance metric \cite{Luo2001} for $D$.

Given an initial parameter set $\{t_0, \cdots, t_N \}$ evenly sampled across $[0, 1]$, we first compute the cumulative rectified arc lengths at each $t_i$ as follows:
$$s\left(t_i\right) = \sum_{k=1}^iD(C(t_i), C(t_{i+1})), \quad \text { for } i=1,2, \ldots, N$$
\\
with the initial condition that:
$$
s(t_0)=0
$$

Dividing each $s\left(t_i\right)$ by the total arc length $s(t_N)$ gives us the relative cumulative arc length at each parameter point:

$$
\hat{s}(t_i)=\frac{s(t_i)}{s(t_N)}
$$
\\
and allows $\hat{s}(t_i)$ and $t_i$ to share the same domain $[0, 1]$. We then approximate $\hat{s}^{-1}$ that satisfies $\hat{s}^{-1}(\hat{s}(t_i)) = t_i$ by inversely mapping each relative cumulative arc length to a parameter point and performing interpolation. 

\begin{figure}[bh]
  \centering
  \vspace{5mm} 
  \includegraphics[width=1\columnwidth]{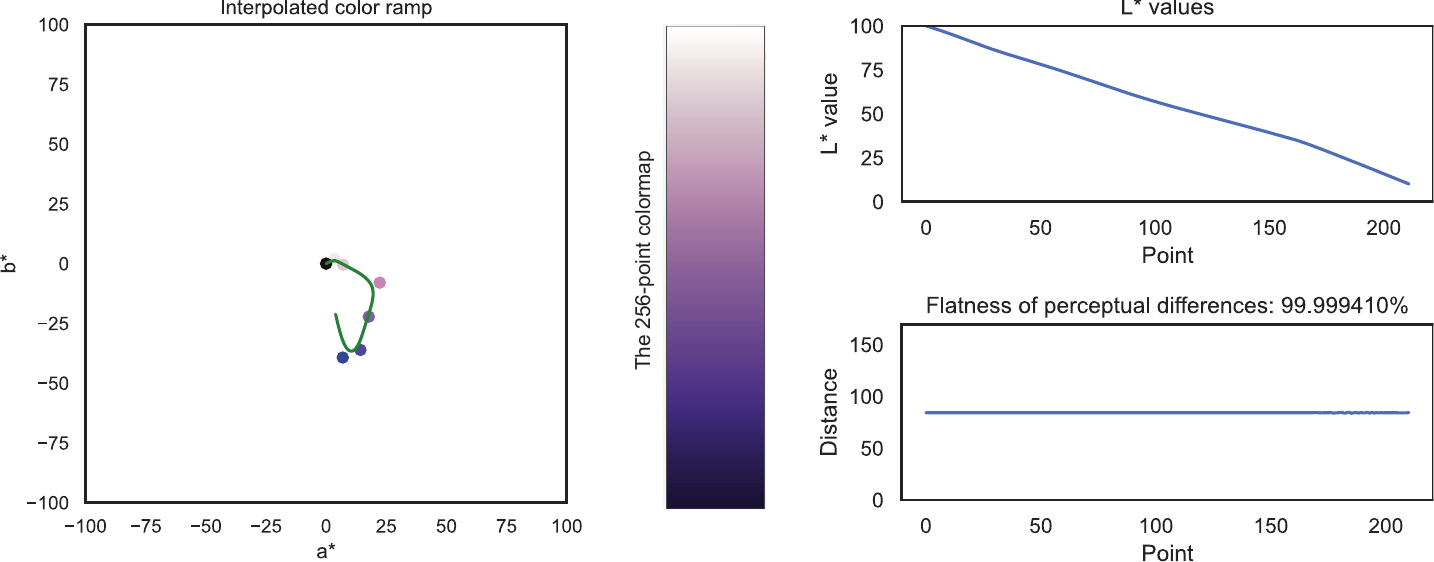}
  \caption{\add{\textbf{Profile of a colormap created by Cieran.} \textmd{Each colormap output by Cieran will be smooth, monotonically varying in lightness and perceptually uniform. The bottom right subplot represents the perceptual distance between each adjacent pair of colors in the 256-point continuous colormap. The ``flatness'' metric is 1.0 minus the std. dev. of distances divided by total arc length.}}}
  \Description[Profile of a colormap created by Cieran.]{Profile of a colormap created by Cieran. On the left is the colormap's projection onto the A-B plane in the CIELAB colorspace, and shows a smooth curve. In the middle is the output colormap. On the left, there are two vertically aligned subfigures. The top subfigure shows the monotonically decreasing trend of the colormap's lightness values. The bottom subfigure shows a completely flat line, indicating the flatness of perceptual differences which is 99.99 percent across each adjacent pair of sampled colors.}
  \label{fig:cmap-profile}
\end{figure}

Finally, we obtain new parameter points by applying the interpolated $\hat{s}^{-1}$ to evenly spaced values within $[0, 1]$. Although the resulting outputs $\{t^{*}_0, \cdots, t^{*}_N \}$ will no longer be evenly spaced within $[0, 1]$, their corresponding coordinates $\{C(t^{*}_0), \ldots, C(t^{*}_N)\}$ will be evenly spaced in terms of rectified arc lengths $D(C(t_i), C(t_{i+1})) = \Delta E_{2000}(C(t_i), C(t_{i+1}))$ (Figure \ref{fig:cmap-profile}).

\section{Additional Qualitative Findings}
\label{appendix-B}
\begin{figure}[bh]
  \centering
  \includegraphics[width=1.0\columnwidth]{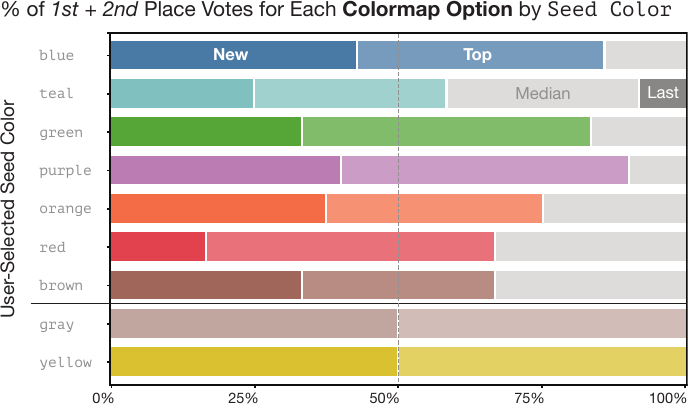}
  \caption{\textbf{Summary of participant rank responses from the user study as a function of their initial seed color choice.} \textmd{Cieran performs well across a variety of seed colors.}}
  \Description[A horizontal 100 percent stacked bar graph where each bar group is a different user seed color]{A horizontal 100 percent stacked bar graph summarizing participant rank responses from the user study. Each bar group is a different initial seed color. For every seed color, most participants ranked new and top colormaps as either first or second.}
  \label{fig:rank-sorted}
\end{figure}

We further explored the robustness of Cieran toward different user-specified seed colors in our Evaluation. Figure \ref{fig:rank-sorted} represents the distribution of 1st and 2nd place votes in the final ranking task in the study. Each bar corresponds to a colormap category (new, top-ranked, median-ranked, last-ranked). We can verify that the majority of 1st and 2nd place votes went to the new and top-ranked colormaps for every seed color.

However, our study comprised 36 total design sessions, and not all colors were selected with equal frequency. Bar groups have been sorted based on the frequency with which that seed color was selected. Although participants that initialized Cieran with the bottom two colors (gray and yellow) never gave 1st and 2nd place votes to median- and last-ranked colormaps, the sample sizes for these two colors are each one. Beginning from the top color (blue), each seed color was selected with the following frequency: 

\begin{itemize}
\item blue: 7 
\item teal: 6 
\item green: 6
\item purple: 5
\item orange: 4
\item red: 3
\item brown: 3
\item gray: 1
\item yellow: 1
\end{itemize}

\end{document}

%% file: tex/introduction.tex
When visualizing data, analysts frequently need to decide what colors to use in a chart.
We focus on the design of sequential colormaps, which must be orderable, appear smooth, include discriminable colors, and adhere to perceptual uniformity \cite{Bujack2018}.
Beyond these perceptual guidelines, colormaps should also have aesthetic appeal \cite{Moreland2009},
which can depend on color-data semantics (e.g., cool--warm, positive--negative affects) \cite{bartram2017affective}, branding, domain conventions, or personal preferences \cite{Palmer2013}.
Furthermore, how we see colors applied on a chart will vary with the sizes and shapes of the colored marks \cite{szafir2017modeling}, their spatial distribution \cite{zimnicki2023effects}, and the chart’s background color \cite{Palmer2013}.
These factors and their interactions make colormap design challenging for a typical data analyst.

Most often, people choose a colormap from a small gallery of options designed by experts like ColorBrewer \cite{Brewer2003} or defaults provided in visualization authoring tools, which greatly limits expressivity.
Even though many other quality example colormaps are available, for example, across communities like ColourLovers,\footnote{https://www.colourlovers.com/} the process of rendering one's visualization for each colormap across galleries and tracking best options can be time-consuming. 

Furthermore, pre-existing colormaps that satisfy user needs may not exist. People can build new colormaps manually using tools like Photoshop; tools like CCC-Tool \cite{Nardini2019} aim to streamline the workflow for visualization.
However, most people lack the technical or design expertise for working with color to prototype and refine colormaps effectively.
This means that without tools that enable data analysts to efficiently \textbf{rank} existing colormaps or \textbf{create} new ones tailored to their data, people will continue to inadequately present data by relying on a system default \cite{moreland2015we}.

We address this issue with Cieran (pronounced KEE-ruhn), an AI assistant that helps analysts rapidly rank and create sequential colormaps when designing communicative visualizations.
In its training phase, Cieran adaptively and iteratively asks
an analyst to choose between two different versions of their visualization each employing a different expert-designed colormap. 
This approach leverages prior work in color science which finds that presenting alternative choices is known to elicit the most reliable human responses for studying color preferences \cite{Palmer2013} while also being an easy task to respond to \cite{biyik2019asking}.
Cieran uses this choice data to update a model of aesthetic utility, which can be used to rank expert-designed sequential colormaps.
It also creates new colormaps by treating their design as a path planning problem across the CIELAB space using the learned model as a reward signal \cite{Sadigh2017}.
All outputs monotonically decrease in L*, are perceptually uniform (Appendix \ref{appendix-A}), and are interpolated using cubic B-splines \cite{DeBoor1972} to meet best colormap design practices.

\textbf{Contributions.} Our main technical contributions (\S\ref{technical}) include 1. an active learning strategy to induce a personalized model of aesthetic utility for colormaps from few inputs, and 2. a fast algorithm to create new colormaps given the utility model.
We validate our technical approaches by conducting a user study (\S\ref{evaluation}) where domain experts ($N=12$) across physical and social sciences took part in an open-ended design task. 
Our findings show that Cieran can efficiently suggest and create aesthetically pleasing colormaps tailored to individual users in around two minutes of use. 

The system implementation of the above approach is a publicly accessible Python package\footnote{https://github.com/matthhong/cieran/} for use within Jupyter Notebooks to rank and create chromatic colormaps according to a user's indicated aesthetic preferences. 
Cieran's workflow is as follows:
\begin{itemize}
\item ask the user for a required (seed) component color,
\item fit expert-designed colormaps to this color and process them to be ordered, smooth, and perceptually uniform,
\item iteratively display a few ($< 15$) pairs of expert-designed colormaps to compare on the user's target visualization,
\item rank all expert-designed colormaps from the comparisons, 
\item create a new processed colormap that matches the user's aesthetic preferences,
\item display a selection tool for the user to choose their colormap from the ranked choices.
\end{itemize}

%% file: tex/background.tex
Automation can support the design of effective visualizations. 
For example, tools like ShowMe \cite{mackinlay2007show} and Draco \cite{moritz2018formalizing} provide visualization recommender systems grounded in known best practices or experimental results. 
Alternatively, visualization linters \cite{chen2021vizlinter,hopkins2020visualint,mcnutt2018linting} 
assist in reviewing and identifying potentially ineffective or misleading design practices. 
These tools emphasize generalizable concepts of perceptual effectiveness (e.g., the ability of a population to accurately conduct visualization tasks within a controlled study \cite{quadri2021}), 
while paying limited attention to residual aspects of design like personal preferences.

In contrast, the human-in-the-loop approach of ViA \cite{healey2008visual} 
incorporates personal preferences into visualization recommendations by asking the user to input their objectives and respond to suggestions. 
However, using ViA requires the user to provide significant scaffolding for the system to work by manually inputting parameters such as importance weights, tasks, spatial frequency, and domain type. 
By focusing on the complex space of colormap design, our work considers how visualization optimization \cite{quadri2023} could be automated with an analyst-in-the-loop by querying them with simple questions, such as: ``Which chart do you like better?''
We ground our work in the science and engineering of colormap design as well as active preference learning.

\subsection{Quantifying Colormap Utility}
Designing useful colormaps remains a fundamental challenge in data visualization.
Bujack et al. \cite{Bujack2018} provided mathematical definitions for three properties of perceptually-effective sequential colormaps: order, discriminative power, and uniformity.
Discriminative power, a quality which aids performance on value comparison tasks \cite{liu2018}, can be improved by varying a colormap's component colors across hue and/or chroma to increase total color variance.

Experts acknowledge that chromatic colormaps should also be aesthetically pleasing \cite{Moreland2009}.
But the aesthetic utility of a colormap will vary with the characteristics of the audience (e.g., personal preference or past experience \cite{ahmad2021red}), the specific dataset (e.g., semantics \cite{schloss2020semantic} and distributions \cite{zimnicki2023effects}), and the problem space (e.g., the target visualization \cite{szafir2017modeling} or the data domain \cite{dasgupta2018effect}). 
For example, how we see colors on a chart varies depending on the sizes, the shapes, and the spatial distribution of the colored marks \cite{szafir2017modeling, zimnicki2023effects}.

Consequences of these
factors and their interactions are so prevalent that tools like Tableau\footnote{https://www.tableau.com/} automatically modify colormaps for different visualization techniques; for example, by decreasing the color lightness values for area charts and choropleth maps \cite{brown2021tableau}.
However, such modifications do not take into account dimensions of aesthetics other than the chart technique used, and quantifying the aesthetic value of a colormap with more specificity remains an open problem in visualization \cite{lau2007towards,he2022beauvis}.

\subsection{Tools for Designing Sequential Colormaps}
To help domain experts customize colormaps for their target context, ColorBrewer \cite{Brewer2003}, matplotlib \cite{Hunter2007} and online communities provide galleries of expert-designed colormaps.
However, searching across hundreds of colormaps for the right look-and-feel can be time-consuming,
and default options may not satisfy user needs.

With tools like Photoshop,\footnote{https://www.adobe.com/products/photoshop/}
people can create new color gradients.
Tools like ColorCAT \cite{Mittelstadt2015} and CCC-Tool \cite{Nardini2019} seek to simplify both colormap creation and validation by allowing users to manually input and adjust control points 
and providing metrics for assessing the output colormap's perceptual utility.
Using such tools requires substantial manual effort and design expertise.
Therefore, AI-assisted tools have leveraged existing examples to guide colormap construction. For example, deep learning can enable colormap style transfer from existing images to new charts \cite{transfer1, transfer2, transfer3, transfer4}.
However, users may not have quality example images. 

Instead, systems can directly leverage the hundreds of expert-designed colormaps as examples.
ColorCrafter \cite{Smart2020} first mines a corpus of 222 colormaps for their structural features and outputs nine model curves that are characteristic of common design practices.
People can apply a seed color to one of the model curves and customize it using affine transforms.
ColorMoves \cite{Samsel2018} allows users to mix-and-match segments of example colormaps.
While such a paradigm may be expressive,
the tool entrusts users with the combinatorial task of mixing and matching colormap segments.

In our work, we aim to 
build a tool that automatically suggests and creates colormaps from examples, in a manner that is driven by users' aesthetic preferences. 
We hypothesize that active preference learning can be used to rank colormaps according to aesthetic utility and automate the assembly of new colormap structures.

\subsection{Preference-Based Design Optimization}
Prior work in computational design have applied active preference learning to enhance the quality of images and renderings.
Their technical contributions often involve algorithmic improvements to Bayesian Optimization from discrete choice data \cite{10.5555/2981562.2981614}, including one-to-many comparisons or learning directly from slider interactions \cite{Koyama2017} or brush strokes \cite{Koyama2022}. Preference learning leverages the power of alternative choice tasks to collect reliable information about the utility \cite{Kahneman1979} of a design artifact using a simple input modality.

Pairwise comparison also elicits the most reliable human responses for studying color preferences \cite{Palmer2013}.
Following prior work, BayesOpt could also be used to optimize certain types of colormaps including Cubehelix \cite{Green2011a}, which parameterizes colormap structures based on a start color, saturation levels, and emphasis on different regions of interest.
However, Cubehelix colormaps have a distinct spiral structure that constrains the space of designs. 

We instead model the aesthetic utility of non-parametric sequential colormaps using a preference-based reward learning (PbRL) algorithm \cite{cheng2011preference}.
PbRL algorithms are typically used to steer the behavior of virtual and physical robots to complement inverse reinforcement learning \cite{abbeel2004apprenticeship, knox2009interactively} or reinforcement learning from human feedback \cite{akrour2012april, thomaz2008teachable}.
We hypothesize they can also guide the design of colormaps, which are continuous and smooth trajectories in a 3D space that 
a steerable robot might traverse.

Cieran uses this intuition to model the aesthetic utility of colormap curves from a few pairwise comparisons to potentially capture a wide range of aesthetic considerations, and in turn, optimize the design of sequential colormaps.

\add{\section{System Objectives}}
\input{tex/objectives-new}

%% file: tex/objectives-new.tex
Quantifying the aesthetic value of colormaps remains an open problem in visualization \cite{lau2007towards,he2022beauvis}. 
Factors that are difficult to generalize such as personal preferences, data semantics, and data patterns can influence the aesthetic appeal of colormaps. 
Although galleries like ColorBrewer, matplotlib's colormap library, and online communities like ColourLovers allow users to explore a range of expert-designed options, manually sifting through colormaps takes significant trial and error.
Furthermore, default options limit expressivity.
Tools like Photoshop or CCC-Tool can support custom colormap creation, but they require considerable manual tuning, along with the time and expertise to effectively navigate this process. Our goal is to simplify the process of customizing sequential colormaps by reducing the need for user expertise and manual cost.
 
Our approach builds on the work by Smart et al. \cite{Smart2020}, which mined structural patterns across existing expert-designed colormap curves.
By incorporating their colormap corpus within a data analysis software, we aim to enable people to \textbf{efficiently rank many example colormaps in the context of the data visualization being designed}, and in the process learn a model of aesthetic utility for colormaps within the user-specific context.
We then aim to use both the learned utility model and the expert examples to \textbf{create new aesthetically pleasing colormaps} to suggest as additional options.
Recent enhancements to CCC-Tool \cite{Nardini2021} include post-processing steps to ensure colormaps meet mathematical formulations of perceptual guidelines outlined by Bujack et al. \cite{Bujack2018}. Their work suggests that tools to create sequential colormaps can and should \textbf{automatically enforce perceptual guidelines} with either constraints or post-processing algorithms.

In summary, we developed a colormap design tool that accomplishes the following objectives:

\begin{itemize}
    \item \textbf{O1}. Allow people to efficiently and effectively sort expert-designed colormaps according to their look-and-feel on the target visualization and given their specific data context.
    \item \textbf{O2}. Create novel colormaps, also driven by expert examples, that provide scenario-specific alternatives to existing options.
    \item \textbf{O3}. Ensure that each colormap satisfies perceptual guidelines for colormap design, such as linear order, smoothness, and perceptual uniformity.
\end{itemize}

%% file: tex/cieran.tex
\begin{figure*}[t]
  \includegraphics[width=\textwidth]{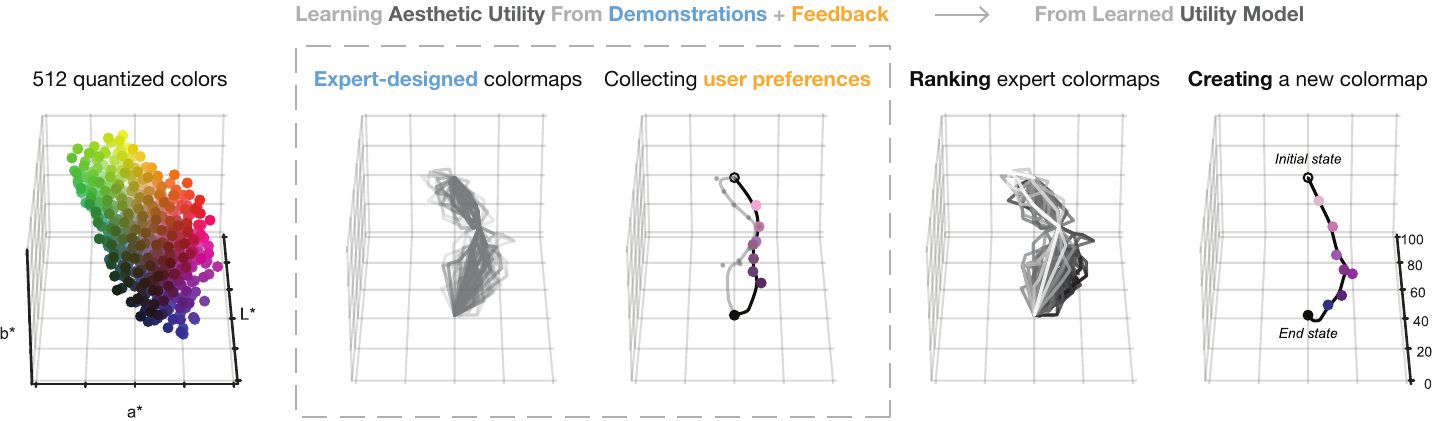}
  \caption{\textbf{Cieran is a path planning agent that helps rank and create sequential colormaps with a human in the loop.} \textmd{
  Cieran first constructs a graphical environment of possible colormap trajectories through a user-specified seed color 
  based on expert demonstrations of colormaps (second subfigure). It then induces a model of aesthetic utility given a small number of pairwise comparison data (the third subfigure highlights the user's selection) collected through a Jupyter Widget, which adaptively and iteratively presents pairs of best candidate colormap examples applied to the user's target dataset. Using the learned utility model, Cieran will score all expert-designed colormaps (fourth subfigure), and create a new quality colormap via path planning (fifth subfigure).}}
  \Description[Five illustrations describing the algorithmic workflow of Cieran.]{Five illustrations describing the algorithmic workflow of Cieran. The first illustration shows 512 colors in the CIELAB gamut. The second illustration shows a set of paths across the gamut from white to black and passing through a user-specified color. The third illustration shows a pair of paths which is displayed to the user for comparison. The fourth illustration shows the original paths again, but they are colored from white to black, where black paths shows the user's preferred paths. The last illustration shows a single path that is generated by Cieran.}
  \label{fig:teaser}
\end{figure*}
\section{Cieran: User Interface}

Cieran is an open-source Python package that interfaces with a user through a Jupyter Widget. This paradigm allows data analysts to both stay within their typical workflow and to design colormaps directly on their target visualization.
Therefore, Cieran can be used to rapidly enhance the quality of any colormapped visualization that can be rendered inside a Jupyter Notebook \emph{in situ}
(\textbf{O1}). 

Using Cieran involves three steps. 
First, the user \textbf{\textcolor{initcolor}{a. initializes}} a Cieran object with a target visualization and color. The system then displays a widget interface where the user \textbf{\textcolor{teachcolor}{b. teaches}} a model their aesthetic preferences for colormaps through pairwise choices (Figure \ref{fig:interface}). Based on these inputs, Cieran allows the user to  \textbf{\textcolor{outputcolor}{c. select}} a colormap from a list of options sorted (Figure \ref{fig:options}) according to their aesthetic utility.

\textbf{\textcolor{initcolor}{a. Initialization.}} Users initialize Cieran with a callback function that accepts a \texttt{matplotlib.colors.ListedColormap} object and displays the target colormapped visualization, in other words, \texttt{Cieran(draw: Callable[[ListedColormap], None])}. Then, the user specifies a seed color that they wish to include across all options with the \texttt{set\_color(color: str)} instance method. The seed color provides an initial entry for the user to indicate their preferences. It also both reduces the number of example colormaps and constrains the search space of new colormaps.

\textbf{\textcolor{teachcolor}{b. Training.}} Cieran is trained by the user iteratively indicating their preference across pairs of example colormaps (Figure \ref{fig:interface}). This training interface is displayed by invoking the \texttt{teach(N: int)} instance method. 
By default, Cieran queries for $N=15$ text responses, with 1 indicating preference for the left colormap option, 2 indicating right, and 0 indicating indifference. 
Pairwise comparison is known to elicit the most efficient and reliable responses for modeling color preferences, compared to rank ordering or absolute value judgments (e.g., a Likert scale response) \cite{Palmer2013}.  

\textbf{\textcolor{outputcolor}{c. Outputs.}} The trained utility model is used to automatically accomplish two design objectives: ranking all example colormap structures (\textbf{O1}) and creating a new colormap (\textbf{O2}) by piecewise combining example colormaps.
Afterwards, the user invokes a slider widget with the \texttt{select()} instance method, where the new and the most useful example colormaps are sorted to the top of a vertical slider (Figure \ref{fig:options}). The colormap selected with the slider can be accessed through the \texttt{cmap} instance property. People can use the \texttt{ListedColormap} object interface 
to further refine the selected colormap (e.g., to resample a smaller number of discrete colors or to represent it as an array of hex values for later re-use).

All colormaps displayed to the user vary monotonically in lightness, and consist of 256 colors sampled at $\Delta E_{2000}$-equidistant rectified arc lengths along an approximate cubic B-spline curve \cite{DeBoor1972} to maintain perceptual uniformity and smoothness (\textbf{O3}, Appendix \ref{appendix-A}).

\section{Cieran\add{: Technical Details}}
\label{technical}

Since colormaps are three-dimensional curves across a colorspace, the problem of finding an aesthetically pleasing colormap is akin to finding a high-utility path within an environment. Cieran executes on this idea by formulating a search space of colormap trajectories, learning a utility model for trajectories from user preferences, and using this model to both rank example colormaps and search for new paths through a colorspace (Figure \ref{fig:teaser}).

\subsection{Technical Overview}

\textbf{Colorspace Quantization.}
Prior to the interpolation of a colormap curve, we define a colormap \textit{trajectory} $\Upsilon = (s_0, a_0, \dots)$ 
as a sequence of control point colors in the CIELAB colorspace \cite{Luo2001} traversed by a virtual \del{robot}\add{agent} based on actions taken across a graph-based environment $\left(S, A\right)$ where:
\begin{itemize}
\item $S$, the state space, is a set of 512 unique CIELAB colors, and
\item $A_s$, the action space of a color $s \in S$, contains possible movements towards neighboring colors.
\end{itemize}

This environment is inherently acyclic (i.e., a DAG), since the agent cannot move backwards in lightness values in order to preserve lightness monotonicity (\textbf{O3}).
Section \ref{sec:quantization} provides details about the construction of this environment.
Our methodology of selecting these 512 approximately equidistant colors in the CIELAB color gamut leads to the mean Euclidean distance between possible control point colors to around $\Delta E=5$, which is the just-noticeable color difference threshold for untrained observers \cite{Mokrzycki}.

\textbf{Learning to Rank.}
Cieran's training phase learns to score each trajectory $\Upsilon$ with
a measure of its aesthetic utility---a user- and context-specific reward function $R(\Upsilon)$---by leveraging a user's pairwise comparison responses.
Assuming perfectly accurate comparisons, precisely ranking $n$ example colormaps would require the user to make $O(n \log n)$ comparisons in the worst case.
Our active learning paradigm adaptively elicits responses to the most informative query pairs to approximate $R$ efficiently from noisy human comparison data.
Section \ref{sec:ranking} provides details about this active learning strategy that can learn $R$ from 15 human labels.

\textbf{Searching for a Novel Colormap.}
Once $R(\Upsilon)$ is learned, we could, in principle, 
use it to score every trajectory in the DAG to find the best novel trajectory (i.e., a combination of piecewise segments from existing colormaps).
However, an algorithm that enumerates and scores all possible combinations of $S$ and $A$ would have exponential time complexity \cite{Tarjan1978}.  
Given a graph, dynamic programming algorithms can achieve path planning to find only the optimal path. 
However, we define the aesthetic utility of a colormap as a holistic quality 
(i.e., the harmony of each color to all other colors), 
not as the sum of utilities across locally adjacent color pairs; such non-additive cost functions preclude the use of combinatorial optimization.
Section \ref{sec:searching} details an approximate dynamic programming algorithm that can still efficiently search for novel colormaps with high cumulative rewards in our graph-based environment.

Cieran forces all colormap trajectories to start from white and end at black to (a) address the potential need for using white to represent zero values and (b) provide matching terminal states across trajectories to simplify the search algorithm. In practice, Cieran's colormap curves (interpolated from the trajectories) are truncated at $L*=10$ by default to exclude the black colors, in accordance with best practices \cite{zheng2022image}.

\subsection{Colormap Quantization}
\label{sec:quantization}
We first define $(S, A)$, the graph-based environment from which Cieran \add{will} sample candidate trajectories through the CIELAB colorspace (Figure \ref{fig:teaser}, first subfigure).
This DAG comprises quantized structures from a corpus of 222 expert-designed colormaps from commercial platforms (e.g., Tableau and R) and communities of practice (e.g., ColourLovers) collated in prior work ColorCrafter \cite{Smart2020}.
Using this corpus constrains the state and action spaces to those reflecting expert designer practices, and quantizing each colormap allows Cieran to create novel colormaps by joining piecewise segments of existing colormaps.

As in ColorCrafter, we first sample nine equidistant points from each expert-designed colormap.
We then align each colormap to a user's seed color by first finding its closest corresponding color with respect to $L^*$.
Then, we generate two alternative options for each of the nine-color ramps.
For the first alternative, we rotate the colormap in the hue plane (i.e., the $a^*$--$b^*$ plane) by the angular difference between the seed and corresponding color.
We then translate the colormap in the $L^*$--$C^*$ plane to complete the alignment.
For the second alternative, only the displacement across the $L^*$--$C^*$ plane between the two colors is used to translate the colormap.

For the state space $S$, Cieran first quantizes CIELAB by generating an approximately equidistant set of 512 colors within the gamut using Halton sampling \cite{Halton1960}, a quasi-random process commonly used in robotics to generate evenly-spaced samples from configuration spaces \cite{Geraerts2004}.
To ensure even dispersion, these positional samples are then passed through the Lloyd-Max algorithm \cite{Lloyd1982, Max1960} until convergence.
For the action space $A$, the nine colors in each new colormap are matched to their nearest colors in the quantized color space.
If a pair of colors $(s, s^{\prime})$ are adjacent in any quantized colormap, we add an action $s^{\prime}$ to $A_s$.
Finally, every lightest color in each colormap becomes the action space of the white state $(100,0,0)$, and every darkest color has an action that corresponds to the black state $(0,0,0)$.

\subsection{Ranking Expert-Designed Colormaps}
\label{sec:ranking}
\input{tex/cieran-ranking.tex}

\subsection{Searching for Novel Colormaps}
\label{sec:searching}
\input{tex/cieran-searching.tex}

\subsection{Implementation}
\label{sec:post-processing}
\input{tex/post-processing.tex}

%% file: tex/cieran-ranking.tex
\begin{figure*}[!htbp]
    \centering
    \includegraphics[width=1\textwidth]{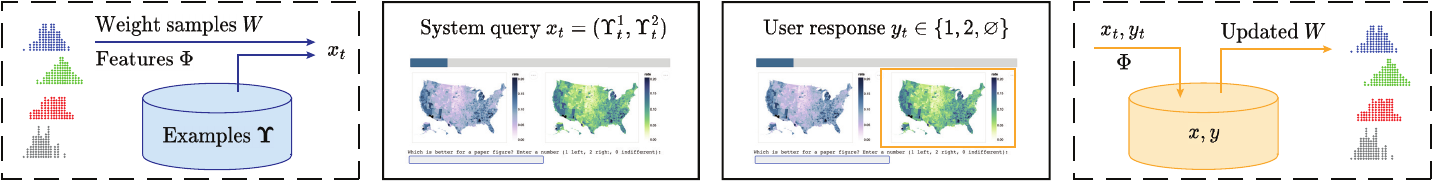}
    \caption{\add{\textbf{Overview of the active learning-to-rank process (Algorithm \ref{preference-learning}).}
    \textmd{[Left] At each iteration, samples $W$ are first drawn from current $P_t(\theta)$ using Metropolis-Hastings. Next, these samples and the feature function $\Phi$ are used to acquire an informative query pair from the example colormap corpus (Eq. 6). [Middle] The user observes this query and makes a pairwise value judgment. [Right] User responses are used to update $P_t(\theta)$ using with the feature function $\Phi$ once again (Eqs. 1 through 5), and the iterative process continues.}}
        }
    \Description[Overview of the active preference learning process]{Overview of the active learning-to-rank process. The first subfigure illustrates how weight samples features are used to generate a pair of example colormaps. In the middle two subfigures, the user evaluates these colormaps and makes a choice. This choice informs the next iterative update of the weight samples, incorporating both the user's feedback and the feature function once again.}
    \label{fig:pref-learning}
  \end{figure*}

For each trajectory $\Upsilon$ through the DAG, we can define a feature vector $\Phi$ that describes its characteristics.
Then $R$, the utility of a trajectory, can be defined as a linear combination these features, such that:
\begin{equation}
    R(\Upsilon)=\theta \cdot \Phi(\Upsilon)
\end{equation}
Algorithm \ref{preference-learning} and Figure \ref{fig:pref-learning} describe the strategy to actively learn $\theta$ from pairwise colormap preferences, which requires the specification of reward features, a belief model, and an acquisition model.

\begin{table}[t]
    \caption{Reward Features and Values}
    \label{tab:params}
    \begin{tabular}{ccl}
    \toprule
    Feature & Description & Value \\
    \midrule
    $\{k_1, \dots, k_8\}$ & perimeter distance & $\mathbf{k} \cdot \theta_{1:8}$ when $s^{\prime}$ = black\\
    $\ell$ & landing reward & 10 when $s^{\prime}$ = black\\
    $m$ & chroma inclination & $m * \theta_{m}$ when $s^{\prime}$ = black\\
    $n$ & moving penalty & $-0.01$ per action\\
    \bottomrule
\end{tabular}
\end{table}

\subsubsection{Reward Model}
Our reward model considers that (1) curves should go through preferred colors (e.g., red) or color families (e.g., warm colors) that a user finds aesthetically pleasing, (2) colormaps should be long enough to be discriminable, but not too long to introduce artifacts like hue banding \cite{bergman1995rule} and (3) that lightness and chroma characteristics are key factors for effective sequential colormaps \cite{Bujack2018}. 
The reward model has nine path features $\{k_1, \dots, k_8\}$ and $m$ (Table \ref{tab:params}) with corresponding unknown weights $\theta_i \in [-1, 1]$, the preference model learned from user choices.

\begin{figure}[t]
    \centering
    \includegraphics[width=1.0\columnwidth]{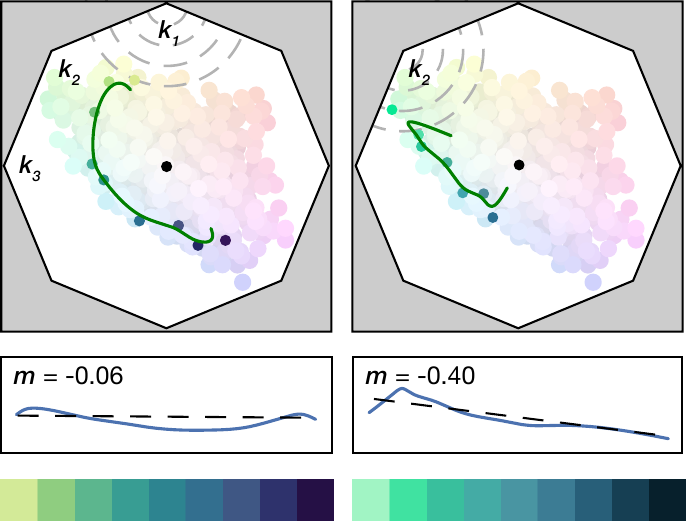}
    \caption{\textbf{A visual explanation of the features and \add{weights} $\theta$ (Table \ref{tab:params})}. \textmd{$\{k_1, \dots, k_8\}$ describes the shortest distance from each of the perimeters in the $a^{*}$--$b^{*}$ plane to the colromap. $m$ is the slope of the colormap in the $L^{*}$--$C^{*}$ plane. These two colormaps represent two trajectories going through the same seed color \#186E8D. The colormap on the left is recommended by Cieran when setting $\theta_1, \theta_2, \theta_3$ to $-0.5$, indicating a preference for a colormap that goes through greenish yellow tones. The colormap on the right is recommended when setting $\theta_2=0.5$, indicating a preference for a colormap that only miminizes distance to the green perimeter of the gamut, while also setting $\theta_{m}=-0.5$, indicating a preference for a colormap that is highly saturated at lower $L*$ values.}}
    \Description[Two-dimensional projections of the color gamut are used to illustrate Cieran's curve features.]{Two-dimensional projections of the color gamut are used to illustrate the curve features. The figure also shows two different paths across the color gamut which significantly in terms of their features.}
    \label{fig:features}
  \end{figure}

Perimeter distances $k$ are the normalized shortest distances between the curve and eight corner points on the boundaries of CIELAB (Figure \ref{fig:features}). 
These distances both reflect preferences for certain color categories and are also indicators of a curve's global length.
Increased curve length may improve the resulting colormap's discriminability \cite{Bujack2018}; however, highly saturate but less preferred hues (i.e., closer proximity to the corresponding perimeter points) are likely to lead to lower aesthetic utility \cite{colorgorical}. 
A negative weight $\theta_k$ indicates that the user prefers a colormap that is close to the color family corresponding to the perimeter $k$.

We model preferences for color chroma (akin to saturation) characteristics as the slope $m$ of each colormap trajectory across the $L^{*}-C^{*}$ plane computed using linear regression.
We normalize the slope such that the colormap with the largest absolute slope has a value of $|m|=1$.
A negative weight $\theta_m$ leads to the lighter colors being relatively more saturated.

\subsubsection{Belief Model}
We initialize the preference model with a non-informative prior by
uniformly sampling points from a unit hypersphere.
Given the conditional independence assumption, we can iteratively update the model as:
\begin{equation}
    P_{t+1}(\theta) \propto P(y_t \mid x_t, \theta) P_{t}(\theta)
\end{equation}
where $x_t=(\Upsilon^1_t, \Upsilon^2_t)$
is the pairwise preference query and $y_t=k$ is the user's response to the query with $k \in \{1, 2, \varnothing\}$, where 1 indicates a preference for the option on the left, 2 on the right, and $\varnothing$ indicating indifference. 

To model a noisy yet rational human expert selecting $k$, we utilize the softmax likelihood function:
\begin{equation}
    \label{eq:softmax}
    P\left(y_t=k \mid x_t, \theta\right)=\frac{\exp (R(\Upsilon^k_t))}{\sum_{\Upsilon \in x_t} \exp (R(\Upsilon))}.
\end{equation}
Some colormap pairs may be equally preferable. 
We therefore include an ``\del{about equal}\add{indifference}'' response \cite{biyik2019asking} by introducing the \emph{minimum perceivable difference threshold} $\delta \geq 0$ \cite{Krishnan1977} such that:
\begin{equation}
    \label{eq:about-equal}
    P\left(y_t=\varnothing \mid x_t, \theta\right)=\left(\exp (2 \delta)-1\right) \prod_{k} P\left(y_t=k \mid x_t, \theta\right) 
\end{equation}
where:
\begin{equation}
    \label{eq:perceivable-difference}
    P\left(y_t=k \mid x_t, \theta\right)=(1+\exp (\delta+R(\Upsilon^{k^{\prime}})-R(\Upsilon^k)))^{-1}
\end{equation}
in which $\Upsilon^k$ and $\Upsilon^{k^{\prime}}$ are the two colormaps being compared and $k \neq k^{\prime}$.
Equation \ref{eq:perceivable-difference} reduces to Equation \ref{eq:softmax} when $\delta=0$.
Prior work shows that the choice of $\delta$ introduces minimal variability, so our algorithm relies on a standard value of $\delta=0.01$ \cite{Biyik2022}. 
Given the above definitions, the Metropolis-Hastings algorithm \cite{Metropolis1953} can adaptively generate samples from the posterior distribution over $\theta$ after obtaining each participant response $k$.

\subsubsection{Acquisition Model}
To sample a pair of colormaps, we use the query-by-disagreement method from Katz et al. \cite{Katz2019}. 
This technique can sample a pair of colormap trajectories that maximize both their potential for aesthetic utility as well as disagreement (i.e., likely differences between colormap utilities).

Let $W$ be samples from the posterior distribution over $\theta$, and let $W_i$ and $W_j$ be the $i$th and $j$th samples of $W$. 
The optimization problem for generating the next query becomes:
\begin{equation}
    \operatorname{argmax}_{i, j} P(\theta=W_i) P(\theta=W_j)+\lambda\left\|W_i-W_j\right\|_2
\end{equation}
where $i \neq j$. 
We can estimate $P(\theta=W_i)$ using Gaussian KDE.
The sampled query pair $x$ will be colormaps whose trajectories $(\Upsilon^1, \Upsilon^2)$ maximize $W_i \cdot \Phi(\Upsilon)$ and $W_j \cdot \Phi(\Upsilon)$ respectively.
$\lambda \geq 0$ is a temperature parameter to incentivize divergent queries; we set it to 500 based on piloting and recommendations from Katz et al \cite{Katz2019}. 

\begin{algorithm}[h]
    \caption{Learning $\theta$ From Colormap Preferences}\label{preference-learning}
    \begin{algorithmic}[1]
        \Require{$\boldsymbol{\Upsilon},\Phi, L$} \Comment{Trajectory set, feature vector, iteration limit}
        \Ensure{$\overline{W}$} \Comment{Parameter estimates $\theta$}
        \State{$t \gets 0, x \gets \varnothing, y \gets \varnothing, W \gets \textsc{MetropolisHastings}(x, y,  \mathbf{0})$}
        \Repeat
        \State {$(\Upsilon^1, \Upsilon^2) \gets \textsc{QueryByDisagreement}(\boldsymbol{\Upsilon}, W, \Phi)$}
        \State {$x \leftarrow x \cup (\Upsilon^1, \Upsilon^2)$}
        \State {$y \leftarrow y \cup \textsc{ObtainUserPreference}(\Upsilon^1, \Upsilon^2) $}
        \State {$W \gets \textsc{MetropolisHastings}(x, y, \overline{W})$}
        \State {$t \gets t + 1$}
        \Until{$t = L$}
    \end{algorithmic}
\end{algorithm}

%% file: tex/cieran-searching.tex
\subsubsection{General Q-Learning Approach}
Cieran utilizes an approximate dynamic programming algorithm known as Q-learning \cite{Watkins1989}, as outlined in Algorithm \ref{q-learning}.
The algorithm observes a reward $r$ after taking each action and estimates $Q$---the value of the action (i.e., appending a color to the current colormap trajectory)---as:
\begin{equation}
\label{eq:q}
Q(s, a) \leftarrow \left(1-\alpha\right)Q\left(s, a\right)+\alpha\left(r+\gamma \max _{a^{\prime}} Q\left(s^{\prime}, a^{\prime}\right)\right).
\end{equation}

Q-learning is typically used to solve a Markov decision process (MDP) where $r$ is defined only in terms of the states and actions, and the utility of a trajectory is the sum of the rewards at each step. 
However, Cieran evaluates colormaps based on their holistic aesthetic utility, and not as a sum of utilities for adjacent pairs of colors.
Therefore, its agent will only receive positive rewards at the last step based on the entire trajectory taken.

Beginning its search from the color white, at each iteration $t$, the algorithm selects an action using an epsilon-greedy strategy and observes a reward.
As described in Table \ref{tab:params}, there is a default cost of $n=-0.01$ to every action to prevent unnecessary movements which do not add to the cumulative reward. This action cost discourages control colors that cause unnecessary chroma or hue variance without contributing to the colormap's aesthetics.

Based on chosen actions and observed rewards, Cieran iteratively updates the state-action value function $Q(s, a)$ using the incremental update rule (Equation \ref{eq:q}) until it reaches the black (end) state.
If the next state is the black state, a landing reward of $\ell=10.0$ is added to the cumulative reward along with $R(\Upsilon)=\theta^{\top} \Phi(\Upsilon)$ based on the features of the final colormap trajectory.

When Cieran reaches the black (end) state,
if the completed trajectory $\Upsilon$ meets the following criteria, it is saved as the new $\Upsilon^{*}$:
\begin{enumerate}
    \item it is novel, as in not already in the corpus (since we have already scored all existing colormaps);
    \item it passes through at least two points between white and black, one of which is the seed color;
    \item every point along its interpolated path is within the gamut;
    \item and it has the highest cumulative reward (i.e., the highest aesthetic utility). 
\end{enumerate}
\subsubsection{Hyperparameters}
Our Q-learning environment uses a non-deterministic transition model where $T(s, a, s') =0.95$ when $a=s^{\prime}$ but transitioning to a randomly different adjacent state with probability $0.05$.
Since our environment is not an MDP typically suited to Q-learning, incorporating additional noise to the reward signal makes the algorithm more robust.
Q-learning requires a discount parameter $\gamma$, but because every trajectory has the same absorbing state (black), $\gamma=1.0$ \cite{Watkins1992}.
Both the exploration parameter and learning rate are set to $\epsilon=0.1$ and $\alpha=0.1$ based on piloting.

We primarily follow standard Q-learning practices except for the initialization of $Q$.
Specifically, the initial action-value estimates $Q_0$ are set to 100 for all actions, which is significantly higher than what can be observed from the reward model.
This Optimistic Q-learning \cite{10.5555/3312046, NIPS2001_6f2688a5} approach encourages early exploration by the model by making all unexplored options appear better than they actually are.
We evaluate the efficacy of this approach in Section \ref{q-evaluation}.

\begin{algorithm}[h]
    \caption{Searching for a Novel Colormap From $\theta$}\label{q-learning}
    \begin{algorithmic}[1]
        \Require{$\theta, \Phi, L, \epsilon, \alpha, Q_0$} \Comment{Preference model, iteration limit, and hyperparameters}
        \Ensure{$\Upsilon^{*}$} \Comment{Found high-utility trajectory}
        \For{$\forall s, a$}
            \State{$Q(s, a) \gets Q_0$}\Comment{Initial optimistic estimate}
        \EndFor
        \State{$t \gets 0, \Upsilon^{*} \gets \varnothing, r^{*} \gets -\infty$}
        \Repeat
            \State{$s \gets white, \Upsilon \gets \{s\}, r \gets 0$}
            \Repeat
                \State {$a, s^{\prime} \gets \textsc{EpsilonGreedy}(\epsilon, Q, s)$} 
                \State {$\Upsilon \gets \Upsilon \cup \{a, s^{\prime}\}$}
                \State {$r \gets \textsc{ComputeReward}(\Upsilon, \theta, \Phi)$}
                \State {$Q(s, a) \gets Q(s, a)+\alpha\left[r+\max _{a^{\prime}} Q\left(s^{\prime}, a^{\prime}\right)-Q(s, a)\right]$}
                \State {$s \gets s^{\prime}$}
            \Until{$s=black$} 
            \If {$r > r^{*}$}
                \State{$\Upsilon^{*}, r^{*} \gets \textsc{UpdateResult}(\Upsilon, r)$}
            \EndIf
        \State{$t \gets t+1$}
        \Until{$t = L$}
    \end{algorithmic}
\end{algorithm}

%% file: tex/post-processing.tex
We implement Algorithm \ref{preference-learning} with modifications to the \texttt{APReL} package \cite{Biyik2022}, which uses \texttt{numpy} \cite{harris2020array} and \texttt{scipy} \cite{2020SciPy-NMeth}.
We implement Algorithm \ref{q-learning} with \texttt{numpy} and \texttt{networkx} \cite{hagberg2008exploring}.
B-spline interpolation of curves and sampling $\Delta E_{2000}$-equidistant colors utilize \texttt{coloraide} \cite{Muse2023}. Each colormap is a \texttt{matplotlib.ListedColormap} \cite{Hunter2007} class instance.
We implement the interface using \texttt{ipywidgets} \cite{jupyter-widgets2023}.

%% file: tex/evaluation.tex
We evaluated whether Cieran can effectively rank and create colormaps that satisfy analysts' design needs given a small set of preference inputs. The study first involved a holistic mixed-methods evaluation of the system with fourteen participants who each designed a set of colormaps using Cieran. We then evaluated our Q-learning implementation (Algorithm 2) using participants' learned preference models.

\subsection{Study Protocol}
\subsubsection{Participants}
Fourteen researchers participated in the study.
The first two participants (E1 and E2) were experts in using colormaps in visualization and served as pilot participants.
Although we do not include their results in our quantitative analysis, our discussions with them informed our study design, and their feedback is included in the qualitative analysis. 

We recruited 12 additional social or physical scientists (P1-P12) from the United States who have all had experience building visualizations in a scripting language such as Python, R, JavaScript, or SPSS.
All participants were native or fluent English speakers over 21 years old.
No participant self-reported color vision deficiency, although this was not part of our exclusion criteria.

\subsubsection{Stimuli}
Participants designed colormaps for a discretely-binned 2D kernel density plot rendered in Seaborn \cite{Waskom2021}.
This plot was accompanied by black discrete histograms of the corresponding marginal distributions to the right (y distribution) and top (x distribution) of the heatmap (Figure \ref{fig:stimuli}).
We chose 2D kernel density plots so that the colormaps would be mapped onto smooth distributions and display multiple distinct layers of color values across the length of the colormap, while also minimizing potentially distracting structural artifacts. 
The histograms allowed participants to verify that every colormap represented the same dataset.

\input{assets/stimuli}

Each participant saw one of three pre-generated datasets.
The datasets were isotropic Gaussian blobs in 2D space generated using \verb|sklearn| designed to ensure that the spatial distribution of data covered most of the display area.
These stimuli also avoided symbolic shapes and hotspots that would pop-out and distract the participant from the global color characteristics. 
We randomly assigned one-third of participants to each dataset condition.

\subsubsection{Procedure}
We conducted a mixed-methods study using a web interface in a hybrid remote/in-person setting. All verbal and physical interactions were recorded through Zoom screenshare. 

We first collected informed consent and basic demographic information. We then asked participants a series of questions about their experiences with visualization and color, including their favorite color. We then introduced the heatmap, describing it as a visualization that allows us
``to aggregate [scatterplot] points, making the distributions easier to see. As it turns out, heatmaps can come in a variety of different colors.''
They were instructed to make a sequence of preferential choices between variants of colormaps mapped onto the heatmap to design a journal paper figure.

Before the formal trials, the participant and the interviewer discussed how the participant would usually go about designing a colormap in the given scenario to elicit baseline strategies for colormap design. Participants then moved on to the study, where we collected both quantitative and qualitative user data.

\subsubsection{Quantitative data collection: }
\label{ranking-task}
Participants made 15 comparisons over three phases (one phase per user-specified seed color), totaling 45 comparisons per participant.
We observed during pilot studies that people tended to lose engagement after 15 comparisons.
The last task in each phase was manually ranking colormaps, and the whole phase was designed to last at most two minutes.

To reduce potential confounds from color pickers, participants specified a seed color from the Tableau10 color palette \cite{Stone2016}, an expert-designed palette of ten discrete colors designed for categorical visualization.  
The first phase used the color in Tableau10 that was closest to the participant's favorite color, but participants were free to choose any color, including a previously used color, in other phases. 
For each pairwise comparison, the participant could indicate their preference for left (1), right (2), or indifferent (0) input using a textbox.
They were instructed to choose 1 or 2 if there was even the slightest preference for one of the colormaps or if they simply disliked one more than the other.

Following each set of 15 comparisons, \add{in place of using the slider selection tool (Figure \ref{fig:options}),} the participant rank-voted four different colormaps: (1) the novel colormap found through search, (2) the top-ranked expert-designed colormap according to its learned aesthetic utility, (3) the median-ranked expert-designed colormap, and (4) the lowest-ranked expert-designed colormap.
The four choices were laid out in a 2x2 grid in a random order. 
Participants input their votes (1--4, with 1 indicating the highest aesthetic utility, and 4 indicating the lowest) using an integer widget.

\subsubsection{Qualitative data collection: }
Participants were encouraged to think aloud and verbally reflect upon their choices at any point during the study.
The experimenter also asked questions during the study, including why the participant made a given preference choice in the first trial of the first block and when participants made an unexpected choice based on their past behavior and available information. 
After the study ended, the participants were asked whether they had any other thoughts during the study before being briefed on the study objective, which was to evaluate the efficacy of a colormap recommendation tool.

\subsubsection{Analysis}

We coded the audio transcripts of the interview for insights into the participants' design objectives, prior experiences with colormap customization, and their experience using Cieran.
The ranked-choice data from the quantitative data collection was analyzed using the Plackett-Luce model \cite{Luce1977} to compare the performance of the four colormap options (new, top-rank, median-rank, and last-rank).

\subsection{Results}
\label{results}
\input{tex/results}

\input{assets/convergence}

\subsection{Post-Study Analysis of Algorithm 2}
\label{q-evaluation}

\input{tex/q-evaluation}

\input{assets/regression}

%% file: assets/stimuli.tex
\begin{figure}[b]
  \centering
  \includegraphics[width=1.0\columnwidth]{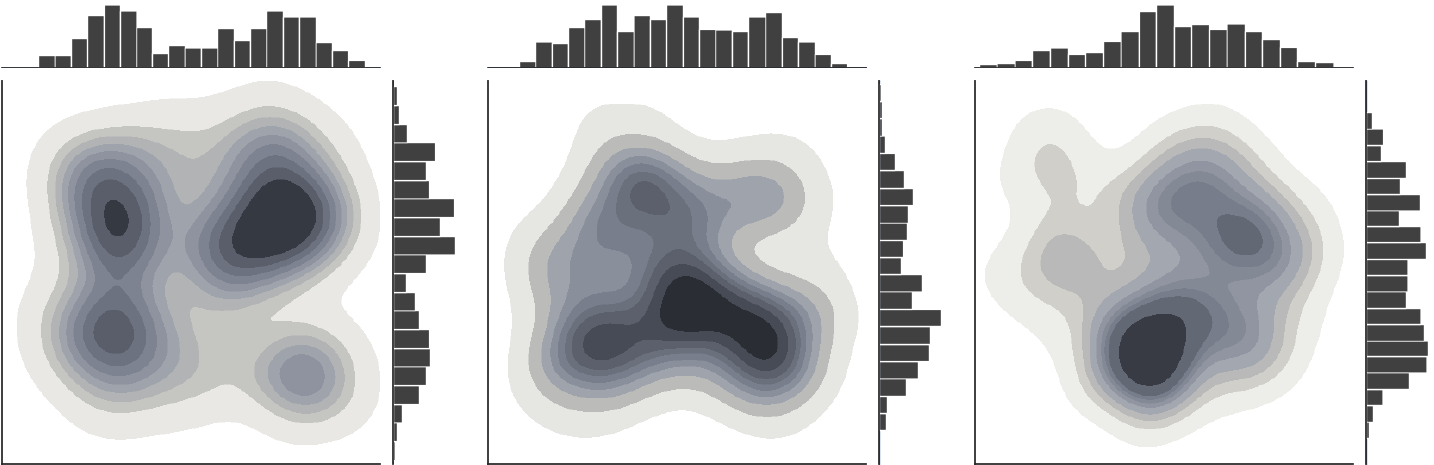}
  \caption{\textbf{The three dataset stimuli used in our study.} \textmd{Participants designed colormaps for a discretely-binned 2D kernel density plot rendered in Seaborn \cite{Waskom2021}.
  This plot was accompanied by black discrete histograms of the corresponding marginal distributions to the right (y distribution) and top (x distribution) of the heatmap.}}
  \Description[Three two-dimensional kernel density plots used as study stimuli]{Three visualization stimuli used in the study. Each stimulus is a two-dimensional kernel density plot rendered in Seaborn.}
  \label{fig:stimuli}
\end{figure}

%% file: tex/results.tex
\input{assets/rank-results}

\subsubsection{Quantitative Results}
\label{sec:quant}
\input{tex/quant}

\subsubsection{Qualitative Results}
\label{sec:qual}
\input{tex/qual}

%% file: assets/rank-results.tex
\begin{figure}[b]
  \centering
  \includegraphics[width=1.0\columnwidth]{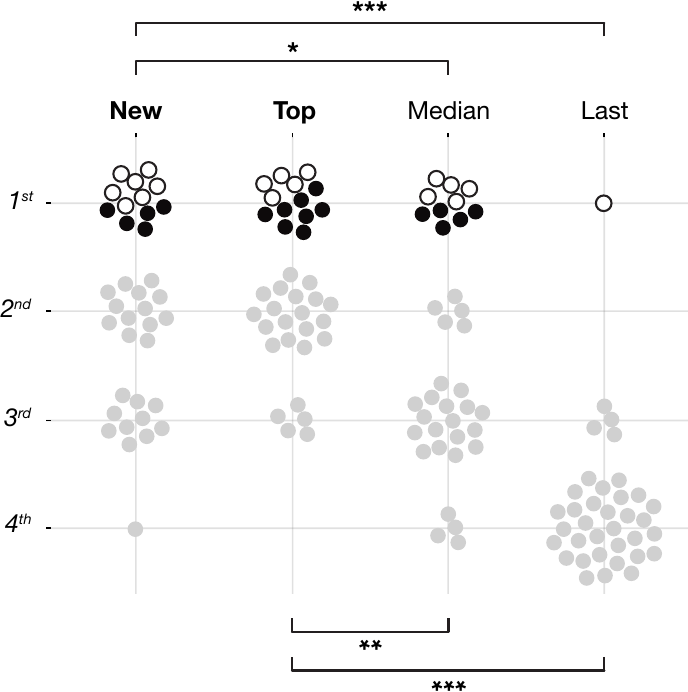}
  \caption{\textbf{Ranked choice votes from the user study.} \textmd{The final task in each phase was to manually rank a shuffled set of four colormaps: Cieran's new colormap and the top-, median-, and last-rank expert colormaps (according to the learned preference model). The new colormap and the top-ranked expert colormap were manually placed higher than others, and in both cases, the differences were significant (* < .05, ** < .01, *** < .001). Of the 36 1st place votes across the top row, filled circles are votes from participants optimizing for colormaps with low hue variance; unfilled circles are votes from participants optimizing for multi-hue colormaps.}}
  \Description[A two-dimensional dot plot showing ranked-choice voting results from the user study]{A two-dimensional dot plot showing ranked-choice voting results from the user study. It shows four colormap options on the X axis and their rankings on the Y axis. Cieran's generated colormaps and its top-ranked colormap consistently have more higher ranked votes than the median-ranked and lowest-ranked colormaps.}
  \label{fig:evaluation}
\end{figure}


%% file: tex/quant.tex
We report inferential statistics, means, and standard errors (means $\pm$ standard errors) for the differences in ranked-choice votes (or, worth\footnote{\add{https://cran.rstudio.com/web/packages/PlackettLuce/vignettes/Overview.html}}) across Cieran's new and example colormaps (Figure \ref{fig:evaluation}).

We did not find a statistically significant difference between new colormaps and the top-ranked expert-designed colormap ($\mu=0.31 \pm 0.30, z=1.05, p=0.30$). However, new colormaps were preferred to the median-ranked ($\mu=-0.67 \pm 0.33, z=-2.02, p<0.05$) and the lowest-ranked ($\mu=-2.75 \pm 0.50, z=-5.42, p<0.001$) expert-designed colormaps.
Similarly, the top-ranked expert-designed colormaps were preferred over the median-ranked ($\mu=-0.98 \pm 0.33, z=-2.96, p<0.01$) and the lowest-ranked ($\mu=-3.06 \pm 0.51, z=-5.98, p<0.001$) expert-designed colormaps.

These results suggest that:
\begin{enumerate}
    \item 
    Cieran effectively ranks colormaps according to aesthetic utility from a small set of personal preference data, and
    \item 
    In aggregate, people found both new and higher-ranked colormaps to be useful, but in some instances (12/36) people may find the new colormap preferable to all expert-designed options, including the top-ranked.
\end{enumerate}

%% file: tex/qual.tex
Our qualitative results reinforced the idea that the aesthetic value of colors is a concept that is difficult to generalize across a population.
Many people indicated an aversion to specific colors, such as yellow (P1, P3, P4), green (P1, P3), pink (P1), and brown (P9).
Some disliked high-chroma colors, such as ``electric” blues, purples, and greens (E2, P1, P8), indicating that they ``hurt [their] eyes” (P8).
However, some specifically preferred yellow (P5), green (P5), ``electric" blue (P6), purple (P7), and pink (P8).

Cieran enabled participants to explore a wide variety of colormaps. 
Many participants ($n=7$) favored ``vibrant” (P3) or multi-hue colormaps (E2, P5) and indicated that high hue variance made data more legible (E2, P2, P3, P5). 
However, such preferences were not universal: many other participants ($n=7$) avoided high variance in hue, worrying that this could distract from communicating data clearly (E1, P1, P10, P2, P4, P12) and might ``overwhelm” people (P12).
Many participants avoided color sequences with highly saturated colors, regardless of whether the colormaps were single-hue (P1, E1, P10) or multi-hue (P7, P8).

Participants who preferred multi-hue colormaps talked about color harmony (E2, P8, P9), color combinations (P5), or how colors ``work together'' (P7) to describe their design objectives.
These participants talked about colormap utility in terms of whether they were ``distinguishable” (P1, P9), ``functional” (P1), or ``crisp” (P3); or, in contrast, whether they were ``blurry”  or ``mushy'' (P3).
They described perceived high-utility colormaps as possessing qualities of ``legibility” (E1, P7) or ``visibility” (P9).
Those who preferred single-hue colormaps seemed to operate with a different objective, often articulating this goal as ``simplicity'' (E1, P10, P2, P4, P12).

The only thing people seemed to agree on was that colors representing lower values should be visible against the white background (E1-E2, P1-P12), so all participants avoided colormaps with minimal variance in chroma.
However, Cieran's ranking and creation capabilities were apparently robust to the variety across participants' design objectives (Figure \ref{fig:evaluation}).
One participant noted that their preferences may vary with their mood (P5), while another noted that they were ``warming up” to certain colors during the study (E2).

%% file: assets/convergence.tex
\begin{figure*}[t]
  \centering
  \includegraphics[width=1\textwidth]{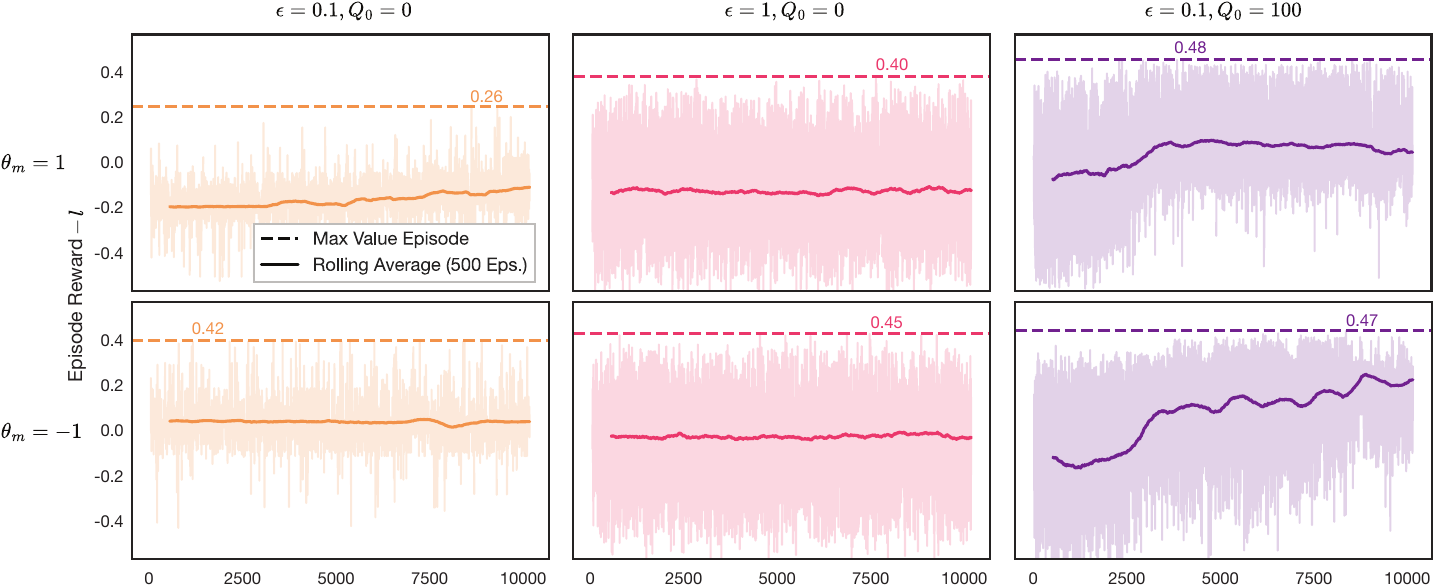}
  \caption{\textbf{Optimistic Q-learning as an adaptive path search algorithm across a DAG.} \textmd{These cumulative reward plots compare the behaviors of three algorithms by depicting the total episode reward (minus landing reward $l$) across 10,000 learning episodes. Each plot depicts a simulation performed with a set of hyperparameters $\epsilon$ and $Q_0$ and with all preference model weights $\theta$ set to $0$ except $\theta_m = 1$ or $\theta_m = -1$ (see Figure \ref{fig:features} for explanation). Traditional Q-learning (left) with $\epsilon = 0.1$ fails to find a high-utility colormap across the DAG, performing significantly worse than random path draws ($\epsilon = 1$, center). However, Optimistic Q-learning (right) with $\epsilon = 0.1$ and $Q_0 = 100$ performs well as a search algorithm by adaptively learning to avoid actions that lead to low-utility colormaps. }}
  \Description[Convergence plots for Q-learning]{Convergence plots for three variants of Q-learning across two different preference models. This chart shows that Cieran's optimistic Q-learning approach outperforms all other approaches by adaptively learning to avoid actions that lead to low-utility colormaps.}
  \label{fig:q-learning}
\end{figure*}

%% file: tex/q-evaluation.tex
Algorithm 2 incorporates a well-known reinforcement learning algorithm to search for a high utility path across the colorspace structured as a directed acyclic graph.
Since the utility of a colormap depends on its entire trajectory, the RL agent receives rewards based on actions taken multiple steps prior; such an environment is not an MDP, and the convergence guarantees typically available for Q-learning~\cite{jaakkola1994convergence} do not apply.
However, this RL agent's goal is not to learn a generalizable model; we incorporated the algorithm within an adaptive search procedure to simply find a useful and novel colormap (Algorithm \ref{q-learning}, line 15) within a reasonable time limit $L$ (which was set to 10,000 trajectory samples for the user study).
In this section, we analyze our agent's behavior to validate that it still improves with experience to better guide its search.

To study the search algorithm and effects of its hyperparameters, we re-ran the Optimistic Q-learning algorithm (with $Q_0=100.0, \epsilon=0.1$) for each participant 10 times using their chosen seed color and the corresponding learned preference model $\theta^{user}$, storing the total reward of the best colormap found after each search. From this total reward we subtracted the landing reward $l$, which is a constant value set at 10 regardless of the user preference model or the algorithm.
We then repeated this for two additional sets of hyperparameters: $Q_0=0.0, \epsilon=1.0$, resulting in a random sampling strategy, and $Q_0=0.0, \epsilon=0.1$, the traditional Q-learning paradigm.
Finally, we employed a linear mixed effects model to quantify the effects of using the three algorithms for colormap search, accounting for the repeated measures on individual models $\theta^{user}$:
$$
\text{maxTotalReward} \sim \text{algorithm} + (1 \mid \theta^{user})
$$

Our findings are summarized in Table \ref{tab:lme}.
On average, the 10,000 iterations of the sampling algorithms took 4.02 seconds on a 2019 MacBook Pro and Python v3.10.12.
Relative to the baseline, traditional Q-learning significantly underperformed ($p < 0.001$) in terms of its output colormap's utility.
However, Cieran's optimistic approach significantly outperformed ($p < 0.001$) other approaches. 

Figure \ref{fig:q-learning} illustrates typical learning behaviors of the path planning agent under the three algorithm conditions.
In the traditional Q-learning paradigm, the agent has a difficult time exploring efficiently within the time limit; the random sampling agent is more likely to have a serendipitous run-in with a high-utility colormap, even though the random sampling agent does not learn and adapt.
On the other hand, Optimistic Q-learning quickly learns to rule out actions that lead to low-utility colormaps by exploring actions more evenly during its early learning stages.
Once the agent's Q-values are updated to more realistic values, it is able to focus on color pairs with higher potential in its search for useful colormaps.

%% file: assets/regression.tex
\begin{table*}[tp]
    \caption{\textbf{Summary of a linear mixed effects model comparing the performance of Q-learning agents to a random sampling baseline.} \textmd{Estimates represent the relative difference from baseline in the cumulative rewards of output colormaps.}}
    \begin{tabular}{lcrlcccccc}
        \toprule
        & \multicolumn{1}{r}{Description} & Estimate & Std. Err. & $z$ & $P>|z|$ & [0.025, 0.975] & \\
        \midrule
        baseline[$Q_0 = 0.0, \epsilon = 1.0$] & \multicolumn{1}{r}{Random Sampling} & 0.550 & 0.115 & 4.789 & < 0.001*** & [0.325, 0.774] & \\
        condition[$Q_0=0.0, \epsilon=0.1$] & \multicolumn{1}{r}{Q-learning} & -0.079 & 0.003 & -30.700 & < 0.001*** & [-0.084, -0.074] & \\
        condition[$Q_0=100.0, \epsilon=0.1$] & \multicolumn{1}{r}{Optimistic Q-learning} & \textbf{0.015} &  0.003 & 5.920 & < 0.001*** & [0.010, 0.020] & \\
        \bottomrule
    \end{tabular}
    \label{tab:lme}
\end{table*}

%% file: tex/discussion.tex
We demonstrated Cieran's capability to effectively rank and create colormaps from user preferences when designing visualizations.
Our system addresses colormap optimization for both aesthetic and perceptual utility by leveraging a corpus of expert-designed curves and processing all output colormaps.
While the prior section summarized these findings given our problem space and the system contribution, we discovered additional observations that might offer insights into the broader scope of research at the intersection of visualization and design optimization.

\subsection{The Diversity of User Preferences}

The diversity of color preferences among participants highlights the individualized nature of aesthetic value.
While we can quantify many aspects of perceptually effective colormap design, taking such a theory-driven approach toward visualization aesthetics may have its limits. 
Instead, Cieran takes an approach that can learn individualized models of aesthetic utility given the analyst's understanding of the intended audience and data context. 

Some aspects of preference may be quantifiable \textit{a priori} with some probability. 
For example, people's backgrounds may impact their preferences. 
Our participants were researchers across the physical and social sciences. 
One expert participant noted that multi-hue colormaps may be preferred by those working in scientific visualization, while single-hue colormaps may be preferred by domain experts working with abstractions of data.
Cultural backgrounds or gender identities may also influence preferences \cite{Palmer2013}.
We may be able to leverage these factors to create foundation preference models that bootstrap optimization processes.
However, they should be applied carefully to avoid perpetuating potential biases, limiting creativity, or transferring designs that call attention to context-specific data semantics to a different setting.

\subsection{Exploring Then Exploiting Preferences}
Some study participants specified their prior preferences for colormaps before using Cieran.
While a few indicated a preference for ``simple'' colormaps that did not vary much in hue,
others noted a prior preference for multi-hue colormaps.
Still, others confessed that they lacked the experience to even have an inclination towards one colormap over the other. 

Yet even those without preferences rapidly identified their own style for colormaps during our study.
P1 and P7 who lacked experience with colormaps at all acknowledged that it was ``very clear how color can work and how it can’t'' (P1) by the end of the study.
P7 described Cieran as being ``insightful and educational for [them], and understanding my taste and preferences for general communication,'' especially when making their pairwise choices.
While P6 did not indicate known prior preferences and had some experience with conventional multi-hue colormaps to visualize astronomical data, they would go on to only select colormaps with low contrast values that they described as ``monochromatic'' or ``subdued'' and indicated that ``simplicity'' was visually appealing to them.

These observations indicate that in addition to reducing the work required to find effective designs, active preference learning tools like Cieran can help people develop a `taste' for what they are seeking in the first place.
This design exploration phase may also be interpreted through the lens of traditional visual data exploration. 
Different colormaps highlight different features in data \cite{tominski2008task,dasgupta2018effect,reda2022rainbow}. 
Using active preference learning tools like Cieran, people can intelligently explore many visualization alternatives, each optimized for highlighting different data patterns.
Although analysts may eventually end up finding a version that looks `just right,' 
we speculate that much as datasets are often represented using multiple chart techniques, analysts may benefit from using multiple colormaps.

\subsection{Diverse Inclinations Toward Agency}
Effective human-AI collaboration must balance automation with agency to effectively leverage the strengths of both people and algorithms \cite{heer2019}. However, this balance may vary across users. 

Participants wanted different levels of control over the output colormaps.
Some participants indicated a lack of confidence in judging the quality of colormapped visualizations.
They also questioned their ability to make internally consistent choices (P3, P6, P9) and worried about their own biases (E2).
These participants expressed a desire for greater automation such as the system making choices for them based on patterns in the data.
However, they still saw Cieran as an improvement on tools that only provided a fixed set of predefined palettes, and noted its efficacy in automating the process of selecting colormaps instead of manually sifting through online galleries of options (P4, P5, P6)

On the other hand, some participants wanted more creative control.
For example, P7 asked to see all ranked options to make a final selection, an available feature in the tool that was not part of the user study (Figure \ref{fig:options}).
Some participants desired an interface to tweak the component colors (P1, E1, P8) or choose the interpolation technique (E1, P8).
The conflict between the desire for more automation (e.g., to overcome perceived time or expertise limitations) and the desire for more control (e.g., to enhance and refine Cieran's outputs) across participants offers insight into how different paradigms for automating visualization design can serve the needs of divergent target user populations.

%% file: tex/limitations.tex
Our system is limited by the number of expert-designed colormaps in the corpus.
Furthermore, displays have a limited color gamut. When
colormaps are rotated and/or translated to fit a seed color, Cieran discards some colormap trajectories because they run outside of the gamut and produce colors that many monitors cannot display.
This gamut constraint motivated us to use the Tableau10 color palette when limiting the selection of seed colors in the user study since colormaps aligned to these colors were
well within the gamut boundaries. 
Although Tableau10 was designed by a visualization expert, we plan to extend the colormap corpus to accommodate as many seed colors as possible.
We also plan to compile a recommended list of `good' seed colors. 

Future iterations of Cieran could consider alternative dimensions of design. At its extreme, the algorithm could leverage techniques that learn feature functions \cite{Katz2019} rather than relying on a prespecified feature set.
However, deep learning approaches will be much slower and require an order of magnitude more comparison data to successfully model user preferences.
P2, P4, and P5 also raised concerns about possible tensions between user preferences and accessibility.
While our approach offers a paradigm that can account for an individual's color vision dynamically, incorporating colorblindness simulations and/or measures to Cieran would allow a more direct path toward addressing accessibility.

Lastly, while Cieran can rank and create variants of both single-hue and multi-hue colormaps effectively, users should be aware of
performance trade-offs between these colormap categories for specific data domain tasks, such as spatial data analysis \cite{dasgupta2018effect} and model checking \cite{Reda2020}. 
Like many visualization construction tools, Cieran assumes that its users have an understanding of their target context, and its algorithms are not guardrails to prevent the construction of potentially inadequate visualizations: it simply acts on user inputs to help users quickly create visualizations that meet universal perceptual guidelines.
Future iterations of Cieran could provide additional scenario-driven support, such as automatically adjusting hue variance based on a set of target tasks or privileging hues that align with domain conventions in addition to seed colors.